\documentclass[reprint,nofootinbib,prd]{revtex4}
\usepackage{eurosym}
\usepackage[utf8]{inputenc}
\usepackage{amsmath,amssymb,amsfonts}
\usepackage{graphicx,float,subfigure}
\usepackage[font={footnotesize,it}]{caption}
\usepackage[colorlinks=true,linkcolor=blue,urlcolor=blue,citecolor=blue]{hyperref}

\begin{document}

\title{Stability of neutral and charged Dyson shells around Reissner-Nordstr%
\"{o}m compact objects}
\author{S. Habib Mazharimousavi}
\email{habib.mazhari@emu.edu.tr}
\affiliation{Department of Physics, Faculty of Arts and Sciences, Eastern Mediterranean
University, Famagusta, North Cyprus via Mersin 10, T\"{u}rkiye}
\date{\today }

\begin{abstract}
In this Letter, we show that, in contrast to Dyson shells surrounding
uncharged compact objects, which are generally unstable, a neutral Dyson
shell enclosing a charged compact object described by the Reissner-Nordström spacetime can attain a stable equilibrium configuration. We
analytically derive the conditions for stability, determine the equilibrium
radius and the corresponding minimum asymptotic energy, and show that small
perturbations about this equilibrium lead to a stable oscillatory motion of
the shell. The oscillation frequency is obtained explicitly and shown to
increase with the shell mass and decrease with the charge of the central
object. When the shell itself carries charge, its stability depends on the
sign of this charge. Shells with the same sign as the central charge become
progressively less stable, while oppositely charged shells exhibit enhanced
stability due to the electrostatic attraction. These findings highlight the
stabilizing role of electromagnetic interactions in Dyson-type thin-shell
configurations within general relativity.
\end{abstract}

\maketitle

\section{Introduction}

In his seminal 1960 paper~\cite{Dyson1960}, F.~J.~Dyson proposed the idea of
a \textit{Dyson shell}, that is a hypothetical spherical structure or swarm of
orbiting collectors surrounding a star and designed to capture a significant
fraction of its radiated energy for use by an advanced civilization. Dyson
argued that such a civilization might enclose its star (or build a swarm of
collectors) and thereby reradiate the absorbed starlight in the mid- or
far-infrared, suggesting that such waste heat could serve as an
observational signature. His short note deliberately avoided engineering
specifics and instead emphasized detectability.

Subsequently, N.~S.~Kardashev proposed a classification of civilizations
based on energy consumption, known as the Kardashev scale (Types~I, II,
and~III)~\cite{Kardashev1964}. A Dyson-scale construct corresponds to a
Type~II civilization, referring to the utilization of stellar-scale energy.
Kardashev's framework stimulated research on the possible observational
signatures of large energy-harnessing structures. Through the late 20th
century, the Dyson concept spread across both technical and popular
literature. However, systematic observational searches began only in the
21st century, when astrophysicists started concrete searches for Dyson-like
waste-heat signatures in archival infrared (IR) data.

Richard Carrigan conducted an IRAS-based whole-sky search and set upper
limits on the number of full or near-complete Dyson shells that could exist
within the catalog~\cite{Richard2009}. This effort established the basis of
the observational program aimed at identifying objects with stellar
luminosities but anomalous mid- or far-infrared colors. Recent reviews~\cite%
{Smith2022} have analyzed the physical feasibility of such megastructures,
discussing the instability of rigid shells, the advantages of swarms,
radiative coupling effects, and potential observational signatures such as
spectral energy distributions and color diagnostics. Wright's review papers~%
\cite{Wright2020} further summarize detectability issues and the limitations
of the idealized \textit{solid-shell} model, favoring instead swarms or
distributed collector systems.

To date, observational searches using IRAS, WISE, and \textit{Spitzer} data
have not produced any unambiguous detections of Dyson-scale waste heat.
Instead, they have placed upper limits on the abundance of complete Dyson
shells or galaxy-spanning civilizations within current survey sensitivities~%
\cite{Richard2009}. Nevertheless, research in this area continues along
three complementary directions: (i)~improved searches using larger infrared
datasets and enhanced photometric precision, (ii)~refined modeling of
engineered infrared signatures accounting for partial coverage, directed
energy use, beaming, and evolutionary effects, and (iii)~engineering and
feasibility analyses that favor distributed swarms over rigid shells.

In a more formal mathematical framework, Berry, Simpson, and Visser~\cite%
{Visser2022} recently modeled a spherical thin-shell megastructure entirely
enclosing a \textit{star-like} central object within general relativity,
thereby formulating a true mathematical analogue of a Dyson shell. They
employed the Israel-Lanczos-Sen junction conditions to match an interior
spacetime to an exterior vacuum, deriving the shell's surface mass density,
surface stresses, and the quasilocal gravitational and mechanical forces
between its hemispheres. The authors examined the classical energy
conditions and found that the Null, Weak, and Strong Energy Conditions (NEC,
WEC, SEC) are satisfied throughout the physically admissible parameter
space, whereas the Dominant Energy Condition (DEC) can be violated for
sufficiently compact configurations, i.e., when the shell approaches the
black-hole threshold. They further investigated the so-called \textit{%
maximum force conjecture}, which posits an upper bound on force in general
relativity, and demonstrated that sufficiently compact shells can violate
the quasilocal version of this bound even when the DEC remains satisfied.
This work thus places the Dyson-shell concept within a rigorous general
relativistic framework, exploring its mechanical and energetic consistency
and identifying the key constraints and potential instabilities that arise
near the black-hole limit.

More recently, Wei \textit{et al.}~\cite{Mann2023} showed that certain
static, spherically symmetric black-hole spacetimes in nonlinear
electrodynamics admit static spherical surfaces at finite radii on which a
massive test particle can remain at rest with respect to an asymptotic
observer. Such static spheres are analogous to classical \textit{Dyson shells%
}, forming equilibrium surfaces that arise purely from the interplay of
gravitational and electromagnetic fields rather than artificial
construction. Using topological arguments, they further demonstrated that,
in asymptotically flat spacetimes, black holes always admit pairs of such
static spheres (one stable and one unstable) when they exist, whereas naked
singularities possess an additional stable sphere. This result broadens the
concept of static spherical shells in general relativity and provides a
theoretical mechanism for their natural occurrence around compact objects,
potentially with observational implications.

Very recently, Hod~\cite{Hod2024} investigated the possibility of
constructing a static, charged thin shell (a Dyson-shell analogue) around a
spherically symmetric compact object in general relativity with Maxwell
fields. Employing the thin-shell junction formalism, he matched an interior
spacetime of mass $M$ to an exterior Reissner-Nordstr\"{o}m region
characterized by the shell's proper mass $m$ and charge $q$, deriving the
conditions under which equilibrium at a finite radius $R$ can be achieved.
By analyzing the total asymptotic energy and imposing both equilibrium ($%
dE/dR=0$) and stability conditions, he found that such configurations are
generally unstable unless finely tuned relations among $M$, $m$, and $q$ are
satisfied (e.g., $q^{2}\simeq m^{2}+2mM$). This result implies that while
Dyson-shell configurations are mathematically consistent within curved
spacetimes, they are highly nongeneric and unlikely to remain stable over
long timescales within standard Einstein-Maxwell theory. 

Motivated by the general notion of Dyson shells in a general relativistic
context, in this Letter we follow the approach of Hod~\cite{Hod2024} and
introduce a Dyson shell surrounding a Reissner-Nordstr\"{o}m compact object
that is dynamically stable. Throughout this work, the Reissner-Nordstr\"{o}m
spacetime with $M<|Q|$ is treated as an idealised theoretical background
within classical Einstein-Maxwell theory. No claim is made regarding the
astrophysical realisability or long-term charge retention of such objects.
The main original contributions of this work are: (i) the demonstration that
a neutral Dyson-type dust shell can achieve stable equilibrium purely due to
the electromagnetic field of a charged central object; (ii) an explicit
analytical derivation of the equilibrium radius via asymptotic energy
minimization; (iii) a closed-form expression for the oscillation frequency
of small radial perturbations; and (iv) a systematic comparison between
same-sign and opposite-sign charged shells within the same formalism.

It should be added that charged thin shell of dust has been studied in the
literature. For instance, de la Cruz and Israel in \cite{Israel1967} studied
the gravitational collapse of a charged thin spherical shell within the
Reissner-Nordstr\"{o}m metric, revealing that under certain conditions the
shell can bounce at a finite radius even after crossing the event horizon.
Although an external observer perceives the collapse as irreversible due to
the event horizon, the shell's proper time description shows it can
re-expand into a distinct, asymptotically flat region of the analytically
extended spacetime. The study highlights the role of electrostatic repulsion
and negative effective mass in enabling such bounces, and discusses
implications for realistic collapse, including the possible lattice
structure of spacetime and the limitations of spherical symmetry and
analytic continuation assumptions. Gao and Lemos in \cite{Gao2007} extended
the results of \cite{Israel1967} to higher dimensions. Kuchar in \cite%
{Kuchar1968} investigates the dynamics of charged spherical shells in
general relativity, focusing on whether electrostatic repulsion can prevent
gravitational collapse. Kuchar generalizes Israel's boundary condition
method to treat charged shells and derives equations of motion that
incorporate surface pressure, charge, and gravitational effects. A key
result is the energy conservation law, which for an observer on the shell
resembles a special-relativistic expression with additional electromagnetic
and gravitational potential terms, while for an external observer it
determines the Schwarzschild mass. The analysis of equilibrium states shows
that shells with low specific charge can be stable under certain conditions,
but shells with high specific charge ($\left\vert e\right\vert \geq M$) can
only be held in unstable equilibrium by pressure, and increasing pressure
inevitably triggers collapse. For dust shells ($p=0$), even when
electromagnetic repulsion dominates ($M<\left\vert e\right\vert$), giving the shell sufficient
energy introduces Nordstr\"{o}m singularities in the exterior field, and the
shell inevitably crosses the upper Nordstr\"{o}m radius in finite proper
time, collapsing from the perspective of an external observer. In \cite%
{Kuchar1968}, it is concluded that no amount of charge can prevent gravitational collapse once
the shell acquires enough energy to generate event horizons. Moreover, Chase
in \cite{Chase1970} derives a general solution for the motion of a charged
spherical fluid shell in a spherically symmetric electrovac field, without
restricting the equation of state. By integrating the equations of motion, a
conservation law for total energy is obtained, generalizing previous results
for dust and polytropic fluids. The stability of equilibrium configurations
is analyzed, showing that for a given equation of state and entropy, there
exists a maximum equilibrium mass. Instability against collapse sets in at a
critical radius that is always larger than the upper Nordstr\"{o}m
gravitational radius, significantly larger for highly relativistic fluids.
Boulware in \cite{Boulware1973} analyzes the collapse of a thin charged
shell and demonstrates that a naked singularity can form if and only if the
shell's proper mass (and thus its matter energy density) is negative. Using
the Israel junction conditions and the maximal analytic extension of the
Reissner-Nordstr\"{o}m metric, Boulware derives the energy equation
relating total mass, proper mass, charge, and radius. For positive proper
mass, collapse inevitably leads to an event horizon shielding any
singularity. Hence, the shell either bounces, collapses through horizons into
other asymptotically flat regions, or reaches $R=0$ only after passing inside
an event horizon. However, if the proper mass is negative, implying
unphysical negative energy density, solutions exist where the shell
collapses directly to $R=0$ without forming an event horizon, producing a
naked singularity observable from infinity. Finally, in \cite{Boulware1973}, it is concluded that proofs
ruling out naked singularities must assume positive matter energy density. As the last paper in this list, we should mention the work of Gao and Lemos in \cite%
{Lemos2021} where the non-collapsing or bouncing charged dust shells in the overcharged regime $M<|Q|$ (the case of our interest) have also been discussed. 

In the works mentioned above, the focus was on the dynamical evolution and bounce of
charged shells in Reissner-Nordstr\"{o}m spacetimes. The present work
differs in scope and emphasis in that we analyze a Dyson-type thin shell
using an asymptotic energy minimization approach, derive explicit
equilibrium radii and oscillation frequencies, and demonstrate that even a
neutral dust shell can attain a stable equilibrium when supported by the
electric field of a charged central object. 

\section{The Dyson shell in a Reissner-Nordstr\"{o}m curved spacetime}

We consider a spherically symmetric thin shell of dust particles with total
proper mass $m$ and radius $R$ formed in the curved Reissner-Nordstr\"{o}m
spacetime. The Dyson shell - in the present context, the term Dyson shell is
used in the idealised general relativistic sense of a thin, spherical dust
shell, following the usage of Berry et al. and Hod - is a spherically
symmetric, timelike, massive thin shell whose interior metric has the line
element (with $G=c=1$) 
\begin{equation}
ds_{i}^{2}=-\left( 1-\frac{2M}{r}+\frac{Q^{2}}{r^{2}}\right) dt^{2}+\left( 1-%
\frac{2M}{r}+\frac{Q^{2}}{r^{2}}\right) ^{-1}dr^{2}+r^{2}\left( d\theta
^{2}+\sin ^{2}\theta \,d\phi ^{2}\right) ,  \label{1}
\end{equation}%
where $M$ and $Q$ denote the ADM mass and charge of the Reissner-Nordstr\"{o}%
m black hole, respectively. The exterior metric represents an \emph{effective%
} Reissner-Nordstr\"{o}m spacetime in the sense that the gravitational
effect of the Dyson shell is included in the line element. Hence, the
exterior metric is given by 
\begin{equation}
ds_{e}^{2}=-\left( 1-\frac{2(M+E(R))}{r}+\frac{Q^{2}}{r^{2}}\right)
dt^{2}+\left( 1-\frac{2(M+E(R))}{r}+\frac{Q^{2}}{r^{2}}\right)
^{-1}dr^{2}+r^{2}\left( d\theta ^{2}+\sin ^{2}\theta \,d\phi ^{2}\right) ,
\label{2}
\end{equation}%
where $E(R)$ is the energy of the shell as measured by an asymptotic
observer~\cite{Hod2024,Hod2013,Hubeny1999}. Applying the Israel junction
conditions \cite{Israel1,Israel2} yields the induced line element on the
shell, 
\begin{equation}
ds_{\Sigma }^{2}=-d\tau ^{2}+R^{2}(\tau )\left( d\theta ^{2}+\sin ^{2}\theta
\,d\phi ^{2}\right) ,  \label{3}
\end{equation}%
and the surface energy-momentum tensor on the shell, 
\begin{equation}
S_{i}^{j}=\mathrm{diag}\left[ -\sigma ,p,p\right] ,  \label{4}
\end{equation}%
where 
\begin{equation}
\sigma =\frac{1}{4\pi R}\left( \sqrt{1-\frac{2M}{R}+\frac{Q^{2}}{R^{2}}+\dot{%
R}^{2}}-\sqrt{1-\frac{2(M+E(R))}{R}+\frac{Q^{2}}{R^{2}}+\dot{R}^{2}}\right) ,
\label{5}
\end{equation}%
and $p$ is determined by the energy conservation condition $S_{i\ ;j}^{j}=0$%
, which gives 
\begin{equation}
p=-\frac{R}{2}\frac{d\sigma }{dR}-\sigma .  \label{6}
\end{equation}

Unlike a generic thin shell, a Dyson shell is composed of dust particles,
for which $p=0$. Therefore, from Eq.~(\ref{6}), one finds 
\begin{equation}
\sigma=\sigma_{0}\left(\frac{R_{0}}{R}\right)^{2},  \label{7}
\end{equation}
where $R_{0}$ is the equilibrium radius satisfying 
\begin{equation}
m=4\pi R_{0}^{2}\sigma_{0}.  \label{8}
\end{equation}
According to Eq.~(\ref{7}), the quantity 
\begin{equation}
m=4\pi R^{2}\sigma  \label{9}
\end{equation}
is conserved, representing the constant proper mass of the shell.

At the equilibrium radius, $\dot{R}=\ddot{R}=0$, and from Eq.~(\ref{5}) we
obtain 
\begin{equation}
\frac{m}{R_{0}}= \sqrt{1-\frac{2M}{R_{0}}+\frac{Q^{2}}{R_{0}^{2}}} -\sqrt{1-%
\frac{2(M+E_{0})}{R_{0}}+\frac{Q^{2}}{R_{0}^{2}}},  \label{10}
\end{equation}
which, upon solving for $E_{0}$, gives 
\begin{equation}
E_{0}=-\frac{m^{2}}{2R_{0}} +m\sqrt{1-\frac{2M}{R_{0}}+\frac{Q^{2}}{R_{0}^{2}%
}}.  \label{11}
\end{equation}
Here, $E_{0}$ consists of two terms, the first represents the average
gravitational self-energy of the shell, while the second corresponds to the
redshifted rest-mass energy of the shell in the curved spacetime of the
central compact object.

Following Hod~\cite{Hod2024}, we determine the radius of a stable Dyson
shell by minimizing the asymptotic energy, that is, 
\begin{equation}
\frac{dE_{0}}{dR_{0}}=0.  \label{12}
\end{equation}
This condition yields 
\begin{equation}
R_{0}=R_{\mathrm{eq}} =M\left( \frac{\left(\frac{2Q}{M}\right)^{2}-\left(%
\frac{m}{M}\right)^{2}} {4-\left(\frac{m}{M}\right)^{2}} \right) \left( 1-%
\frac{m}{M} \sqrt{\frac{\left(\frac{Q}{M}\right)^{2}-1} {\left(\frac{2Q}{M}%
\right)^{2}-\left(\frac{m}{M}\right)^{2}}} \right).  \label{13}
\end{equation}

\begin{figure}[tb]
\includegraphics[width=100mm]{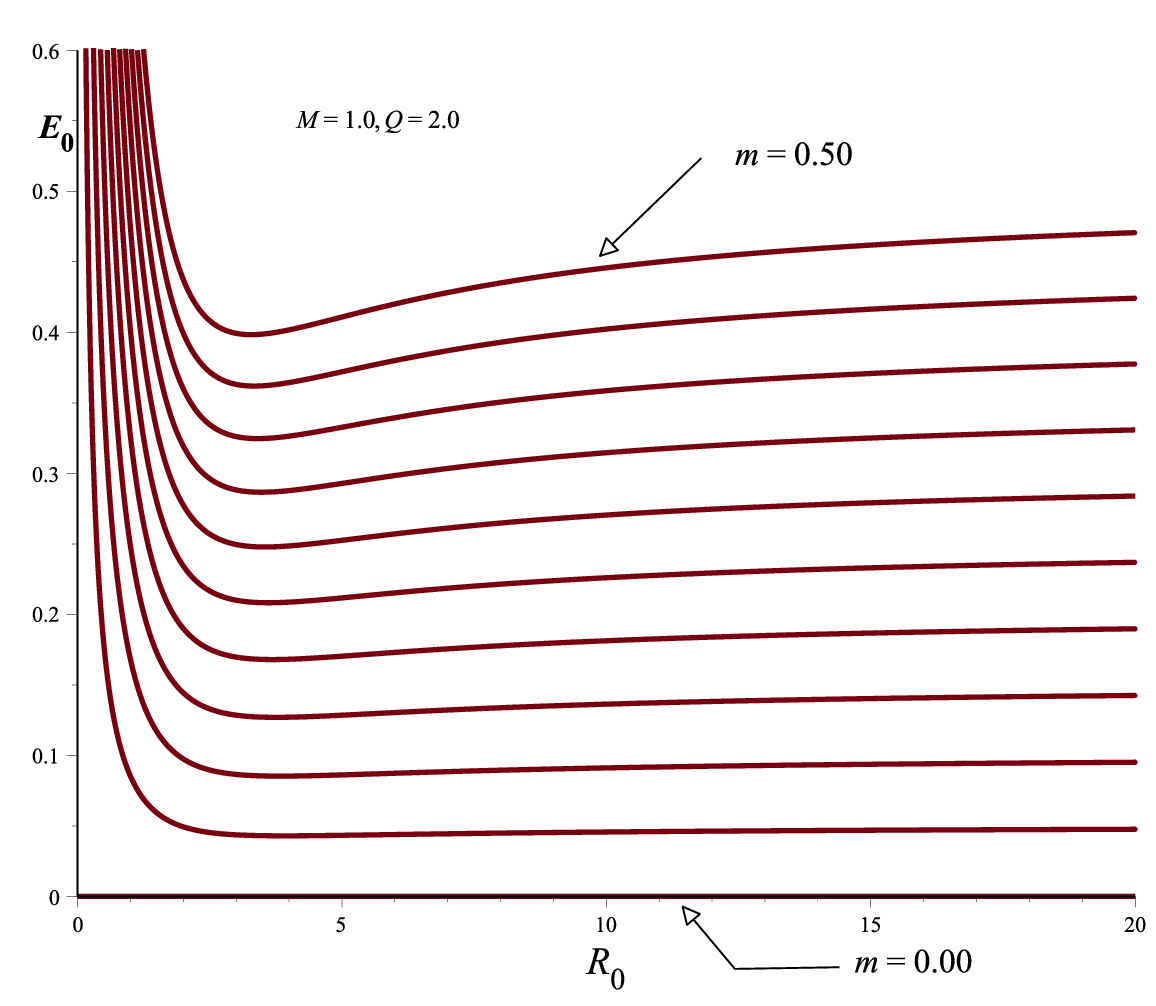}
\caption{The asymptotic energy $E_{0}$ of Eq.~(\protect\ref{11}) as a
function of $R_{0}$ for $M=1.0$, $Q=2.0$, and $m$ ranging from $0.00$
(bottom) to $0.50$ (top) in equal intervals. The minimum energy $E_{0}(%
\mathrm{min})$ corresponds to the stable configuration of the Dyson shell.
Remarkably, the shell remains dynamically stable even in the absence of any
shell charge.}
\label{F1}
\end{figure}

\begin{figure}[tb]
\includegraphics[width=100mm]{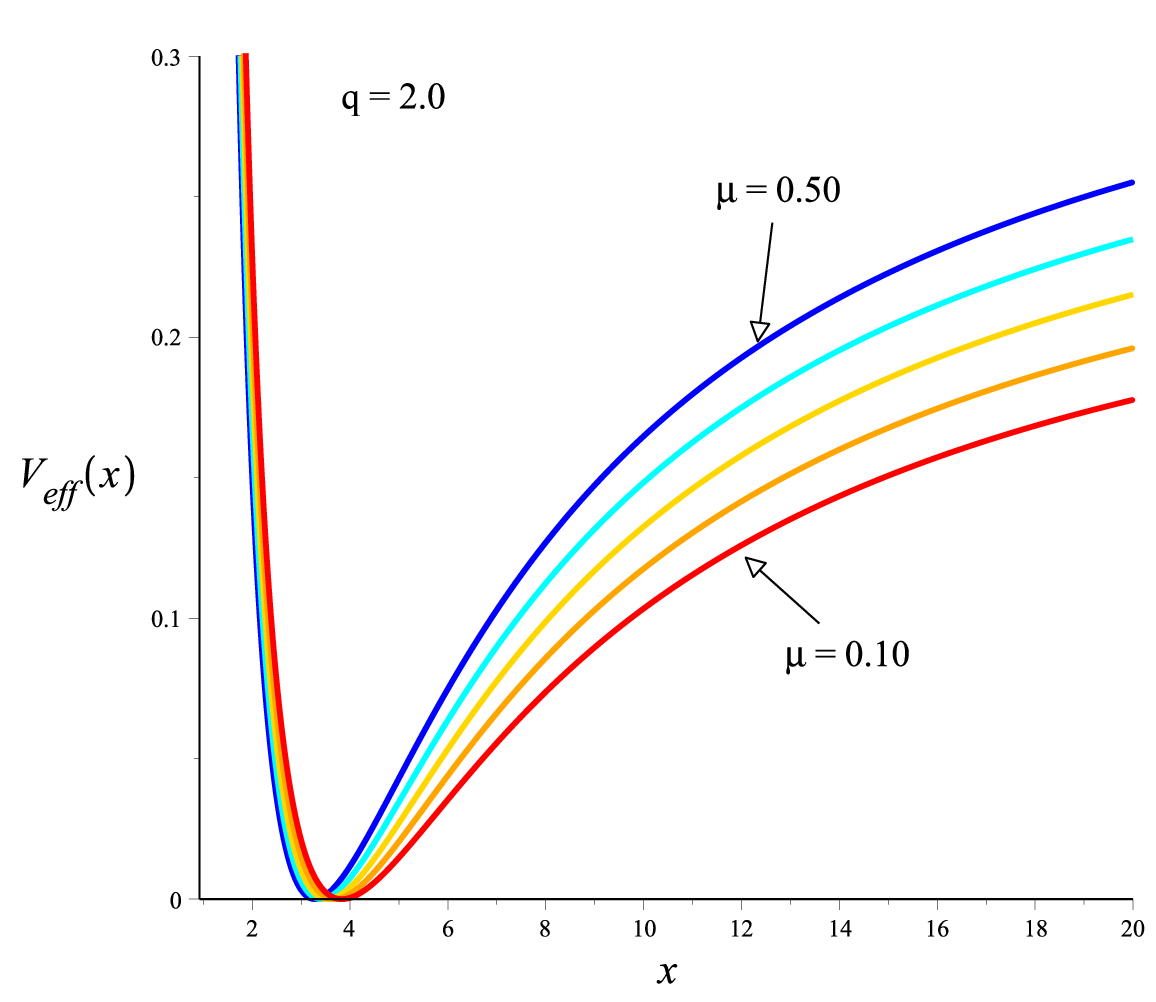}
\caption{The effective potential $V_{\mathrm{eff}}$ of the one-dimensional
motion of the Dyson shell versus $x=R/M$ for $q=2$ and $\protect\mu=0.10$
(bottom) to $0.50$ (top) in equal intervals. The minimum potential
corresponds to the stable equilibrium radius depicted in Fig.~\protect\ref%
{F1}. Remarkably, the shell remains dynamically stable even in the absence
of any shell charge.}
\label{F2}
\end{figure}

\begin{figure}[tb]
\includegraphics[width=100mm]{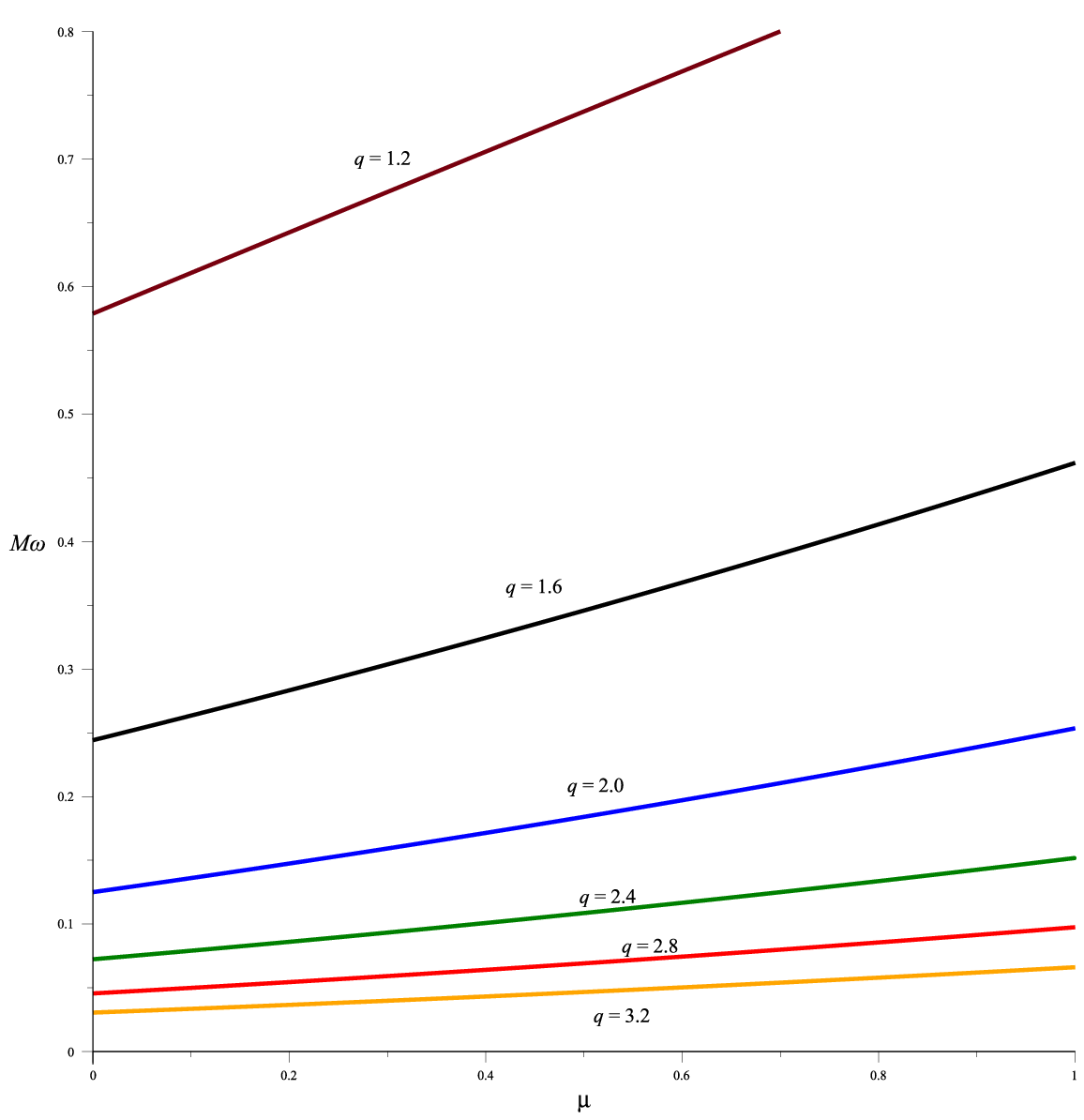}
\caption{The oscillation frequency $\protect\omega$ of the Dyson shell under
small perturbations around the stable equilibrium radius $R=R_{\mathrm{eq}}$%
, plotted versus $\protect\mu$ for $q=1.20$ (top) to $3.20$ (bottom) in
equal intervals. Increasing the shell mass (with fixed $M$ and $Q$)
increases the frequency, indicating stronger stability, while increasing the
charge of the central object (with fixed $M$ and $m$) reduces it.
Remarkably, the shell remains dynamically stable even in the absence of any
shell charge.}
\label{F3}
\end{figure}

Considering the physically relevant configuration in which the mass of the
Dyson shell is much smaller than that of the central black hole or star, we
impose $m<M$. For the equilibrium radius $R_{0}$ to be real, one must assume
either (i) $m<2|Q|$ and $M<|Q|$, or (ii) $m>2|Q|$ and $M>|Q|$. In the second
scenario, the equilibrium radius lies inside the event horizon and is
therefore unphysical. Hence, only case (i) is physically admissible, and we
thus consider $m<M<|Q|$.

In Fig.~\ref{F1}, we plot the asymptotic energy from Eq.~(\ref{11}) as a
function of $R_{0}$ for $M=1$, $Q=2$, and $m$ ranging from $0.0$ to $0.50$.
We observe an attractive potential that yields a stable radius $R_{0}=R_{%
\mathrm{eq}}$, corresponding to the minimum of the energy curve. In terms of
the physical parameters, this minimum energy is given by 
\begin{equation}
E_{0(\mathrm{min})} =M\frac{\mu\left( 4\sqrt{2}\sqrt{(q^{2}-1)\left(2q^{2}+%
\mu^{2}\left(\frac{q^{2}}{2}-1\right) -\mu\sqrt{q^{2}-1}\sqrt{4q^{2}-\mu^{2}}%
\right)} -\mu(4-\mu^{2}) \right)} {8q^{2}-2\mu^{2}-2\mu\sqrt{q^{2}-1}\sqrt{%
4q^{2}-\mu^{2}}},  \label{14}
\end{equation}
where $\mu=m/M<1$ and $q=Q/M>1$.

A stable shell oscillates about $R=R_{\mathrm{eq}}$ when subjected to small
perturbations. Using Eq.~(\ref{5}), we can write the equation of motion as 
\begin{equation}
\dot{R}^{2}+V_{\mathrm{eff}}(R)=0,  \label{15}
\end{equation}
where 
\begin{equation}
V_{\mathrm{eff}}(R)=1-\frac{E_{0(\mathrm{min})}^{2}}{m^{2}} -\frac{E_{0(%
\mathrm{min})}+2M}{R} +\frac{4Q^{2}-m^{2}}{4R^{2}},  \label{16}
\end{equation}
is the effective potential governing the one-dimensional motion of the
shell. Figure~\ref{F2} shows $V_{\mathrm{eff}}$ as a function of $x=R/M$ for 
$q=Q/M=2.0$ and various $\mu=m/M$. The minimum potential corresponds to the
stable equilibrium radius and coincides with the minimum of the asymptotic
energy in Fig.~\ref{F1}.

Expanding $V_{\mathrm{eff}}(R)$ about $R=R_{\mathrm{eq}}$, Eq.~(\ref{15})
becomes 
\begin{equation}
\dot{R}^{2}+\frac{1}{2}V^{\prime\prime}(R_{\mathrm{eq}})(R-R_{\mathrm{eq}%
})^{2} +\mathcal{O}\!\left((R-R_{\mathrm{eq}})^{3}\right)=0.  \label{17}
\end{equation}
At equilibrium, $V(R_{\mathrm{eq}})=V^{\prime}(R_{\mathrm{eq}})=0$ and $%
V^{\prime\prime}(R_{\mathrm{eq}})>0$ (see Fig.~\ref{F2}). Introducing $%
X=R-R_{\mathrm{eq}}$, Eq.~(\ref{17}) becomes 
\begin{equation}
\ddot{X}+\omega^{2}X\simeq 0,  \label{18}
\end{equation}
where 
\begin{equation}
\omega=\sqrt{\frac{1}{2}V^{\prime\prime}(R_{\mathrm{eq}})}  \label{19}
\end{equation}
is the oscillation frequency. Using the dimensionless parameters $\mu=m/M<1$
and $q=Q/M>1$, we obtain 
\begin{equation}
\omega^{2} =\frac{(4-\mu^{2})^{3}\left[ \mu^{2}(4+\mu^{2})
+4q^{2}(4-3\mu^{2}) +8\mu\sqrt{q^{2}-1}\left( \sqrt{4q^{2}-\mu^{2}} -\sqrt{%
q^{2}(4+\mu^{2})-2\mu^{2}-2\mu\sqrt{q^{2}-1}\sqrt{4q^{2}-\mu^{2}}} \right) %
\right]} {4M^{2}\left(4q^{2}-\mu^{2}-\mu\sqrt{q^{2}-1}\sqrt{4q^{2}-\mu^{2}}%
\right)^{4}}.  \label{20}
\end{equation}

In Fig.~\ref{F3}, we plot $M\omega $ as a function of $\mu $ for various
values of $q$. For fixed $M$ and $Q$, increasing the shell mass enhances the
frequency, indicating stronger stability. Conversely, for fixed $M$ and $m$,
increasing the central object's charge $Q$ reduces the oscillation
frequency, implying weaker stability. 

\section{The charged Dyson shell}

\begin{figure}[tb]
\includegraphics[width=100mm]{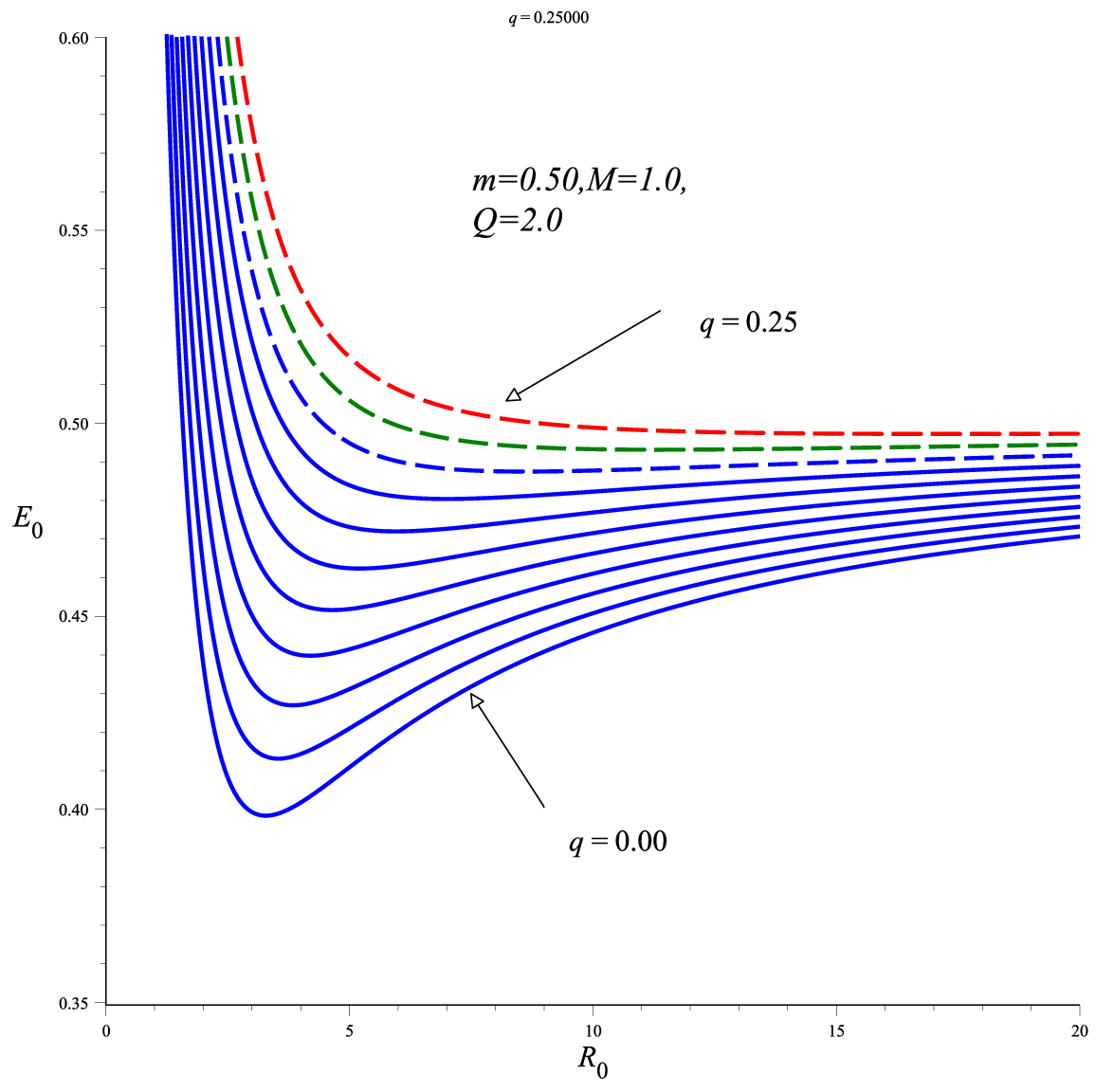}
\caption{The asymptotic energy $E_{0}$ of the charged Dyson shell as a
function of $R_{0}$ for fixed values $M=1.0$, $Q=2.0$, and $m=0.50$, and for
various shell charges $q$ ranging from $0.00$ (bottom) to $0.25$ (top). As
seen, while the neutral Dyson shell is strongly stable, adding charge to the
shell first weakens its stability and eventually renders it unstable.}
\label{F4}
\end{figure}

\begin{figure}[tb]
\includegraphics[width=100mm]{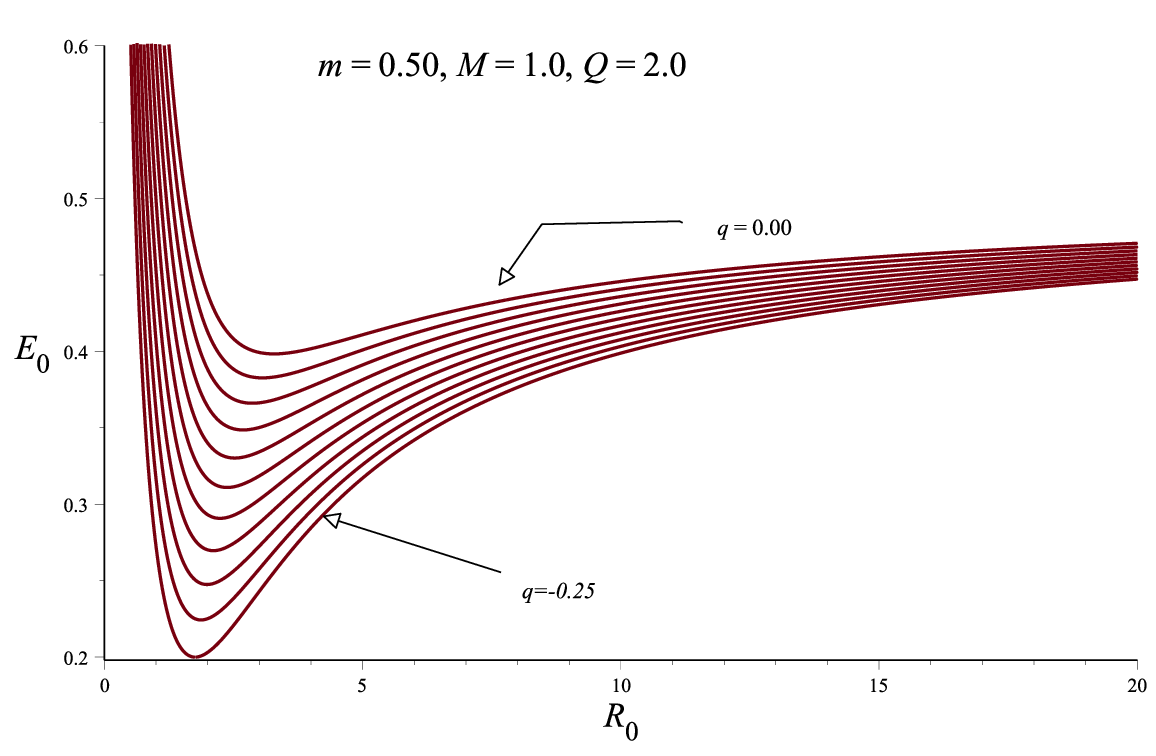}
\caption{The asymptotic energy $E_{0}$ of the charged Dyson shell as a
function of $R_{0}$ for fixed values $M=1.0$, $Q=2.0$, and $m=0.50$, and for
various shell charges $q$ ranging from $-0.25$ (bottom) to $0.00$ (top). The
opposite signs of charge between the central compact object and the shell
enhance the stability of the Dyson shell.}
\label{F5}
\end{figure}

In the previous section, we introduced a stable, spherically symmetric, 
\emph{neutral} Dyson shell composed of dust surroundi Reissner-Nordstr\"{o}m compact object. In this section, we extend the analysis to include 
\emph{charged} dust particles forming a charged Dyson shell in the same
Reissner-Nordstr\"{o}m spacetime. The stability of charged thin shells has
been investigated in different physical contexts; see, for example, Ref. 
\cite{TS1} and the references therein. The interior geometry remains
identical to Eq.~(\ref{1}), while the exterior spacetime is now modified to 
\begin{equation}
ds_{e}^{2}=-\left( 1-\frac{2(M+E(R))}{r}+\frac{(Q+q)^{2}}{r^{2}}\right)
dt^{2}+\left( 1-\frac{2(M+E(R))}{r}+\frac{(Q+q)^{2}}{r^{2}}\right)
^{-1}dr^{2}+r^{2}\left( d\theta ^{2}+\sin ^{2}\theta \,d\phi ^{2}\right) ,
\label{21}
\end{equation}%
where $M$ and $Q$ denote the ADM mass and charge of the central compact
object, and $q$ is the total charge of the Dyson shell.

Following the same formalism as in the neutral case, we find 
\begin{equation}
\frac{m}{R} =\sqrt{1-\frac{2M}{R}+\frac{Q^{2}}{R^{2}}+\dot{R}^{2}} - \sqrt{1-%
\frac{2(M+E(R))}{R}+\frac{(Q+q)^{2}}{R^{2}}+\dot{R}^{2}},  \label{22}
\end{equation}
so that, at an equilibrium radius $R=R_{0}$ with $\dot{R}=\ddot{R}=0$, the
asymptotic energy of the shell becomes 
\begin{equation}
E_{0}=-\frac{m^{2}}{2R_{0}}+\frac{q^{2}}{2R_{0}}+\frac{qQ}{R_{0}} +m\sqrt{1-%
\frac{2M}{R_{0}}+\frac{Q^{2}}{R_{0}^{2}}}.  \label{23}
\end{equation}
Here, the first term represents the average gravitational self-energy of the
shell, the second term is the average electrostatic self-energy, the third
term corresponds to the electrostatic interaction energy between the shell
and the central compact object, and the final term denotes the redshifted
rest-mass energy of the shell.

To determine possible equilibrium configurations, we solve $\frac{dE_{0}}{%
dR_{0}}=0$ to identify the radius $R_{0}=R_{\mathrm{eq}}$ at which the
asymptotic energy reaches its minimum, corresponding to a dynamically stable
configuration.

The explicit calculation yields the equilibrium radius 
\begin{equation}
R_{\mathrm{eq}}= \frac{M(4m^{2}Q^{2}-k^{4})}{4m^{2}M^{2}-k^{4}} -\frac{k^{2}%
\sqrt{(4m^{2}Q^{2}-k^{4})(Q^{2}-M^{2})}} {4m^{2}M^{2}-k^{4}},  \label{24}
\end{equation}
where 
\begin{equation}
k^{2}=m^{2}-q^{2}-2qQ.  \label{26}
\end{equation}
As before, we impose the conditions $M<|Q|$ and $\frac{k^{2}}{m}<2|Q|$ to
ensure that $R_{\mathrm{eq}}$ is real and lies outside the event horizon.

Figure~\ref{F4} illustrates the behavior of the asymptotic energy $E_{0}$ as
a function of $R_{0}$ for fixed $M$, $Q$, and $m$, and for several values of
the shell charge $q$. The results show that when the shell and the central
object carry \emph{charges of the same sign}, the electrostatic repulsion
reduces the depth of the potential well associated with $E_{0}$, thereby
weakening the stability of the configuration. Beyond a critical charge, the
minimum disappears and the Dyson shell becomes unstable. Conversely, when
the charges are of \emph{opposite sign}, the electrostatic attraction
strengthens the binding of the shell and enhances its stability. This
behavior is depicted in Fig.~\ref{F5}, which demonstrates that oppositely
charged shells exhibit a deeper potential minimum and hence a more robust
equilibrium state.

\subsection*{Relation to Cosmic Censorship}

The results presented in this work are obtained within classical
Einstein-Maxwell theory and rely on the use of an overcharged
Reissner-Nordstr\"{o}m spacetime with $|Q|>M$ as an idealized background
geometry. It is therefore important to clarify the relation of these
configurations to the cosmic censorship conjecture. Cosmic censorship, in
its weak form, posits that spacetime singularities arising from
gravitational collapse should be hidden behind event horizons. However,
despite decades of study, the conjecture remains unproven, and several
thought experiments have demonstrated that overcharging or overspinning a
black hole may be possible under carefully controlled conditions \cite%
{Hubeny1999,Hod2013}. In the present analysis, we do not address the
dynamical formation of an overcharged Reissner-Nordstr\"{o}m spacetime, nor
do we claim that such configurations can arise from generic astrophysical
processes. Instead, the overcharged spacetime is treated as a fixed
theoretical background, serving as a laboratory for exploring the mechanical
and dynamical properties of thin-shell structures within general relativity.
In this sense, our results concern the \emph{existence and stability} of
Dyson-type thin shells once such a background is assumed, rather than the
physical feasibility of producing it. It is also important to emphasize that
dynamical stability of a shell at a finite radius does not imply that the
underlying spacetime can be formed or maintained in a realistic setting.
Stability, as analyzed here through energy minimization and small radial
perturbations, should therefore be understood as a local property of the
shell configuration and not as evidence for the violation or evasion of
cosmic censorship. The present work thus complements, rather than
challenges, existing studies on cosmic censorship by focusing on equilibrium
and stability properties in idealized Einstein-Maxwell spacetimes.

\section{Conclusion}

In this Letter, we have shown that a neutral Dyson shell surrounding a
charged Reissner-Nordstr\"{o}m compact object can achieve a stable
equilibrium configuration, unlike previously studied neutral or charged
shells around uncharged objects, which are generically unstable. The
electric field of the central object introduces an additional repulsive
component that counteracts gravitational attraction, giving rise to a stable
equilibrium radius corresponding to the minimum of the asymptotic energy.
Small perturbations about this radius result in harmonic oscillations with
frequency $\omega $, confirming dynamical stability. When the shell itself
is charged, the sign of its charge becomes crucial. A shell carrying the 
\emph{same sign} as the central charge experiences enhanced repulsion and
reduced stability, while an \emph{oppositely charged} shell exhibits
stronger binding and deeper potential minima. These results demonstrate that
electromagnetic interactions can naturally stabilize or destabilize
Dyson-type thin-shell structures in curved spacetimes, providing a novel
theoretical mechanism for equilibrium configurations of matter in
Einstein-Maxwell systems and a foundation for future studies involving
nonlinear or quantum extensions.

\end{document}